\documentclass[12pt]{iopart}
\usepackage{graphicx}
\begin{document}

\title{Generalized information entropies depending only on the probability distribution }

\author{O Obreg\'on and A Gil-Villegas}

\address{Divisi\'{o}n de Ciencias e Ingenier\'{\i}as,
Campus Le\'{o}n, Universidad de Guanajuato, Loma del Bosque No. 103, Fracc. Lomas
del Campestre, Le\'{o}n, Guanajuato, 37150, M\'{e}xico}
\eads{\mailto{octavio@fisica.ugto.mx}, \mailto{gil@fisica.ugto.mx}}
\begin{abstract}
Systems with a long-term stationary state that possess as a spatio-temporally fluctuation
 quantity $\beta$ can be described by a superposition of several statistics, a "superstatistics".
 We consider first, the Gamma, log-normal and $F$-distributions of $\beta$.  It is  assumed
  that they depend only on $p_l$, the probability associated
   with the microscopic configuration of the system.  For each of the three $\beta-$distributions we calculate
   the Boltzmann factors and show that they coincide
   for small variance of the fluctuations.  For the Gamma distribution it is possible to calculate
the entropy in a closed form, depending on $p_l$, and to obtain then an equation relating $p_l$ with $\beta E_l$. We
also propose, as other examples, new entropies close related with the Kaniadakis and two possible
Sharma-Mittal entropies.  The entropies presented in this work do not depend on a constant
parameter $q$  but on $p_l$.

For the $p_l$-Gamma distribution and its corresponding $B_{p_l}(E)$ Boltzmann factor and the
associated entropy, we show the validity of the saddle-point approximation.  We also
briefly discuss the generalization of one of the four Khinchin axioms to get this proposed
entropy.
\end{abstract}

\maketitle

\section{Introduction}
Several measures of information have been proposed in the literature \cite{1}, appart
from Shannon entropy \cite{shannon}. By maximizing these
information measures \cite{2}, their corresponding probability distributions
can be calculated.  Some of these generalized information and entropy measures
and their potential physical applications have been discussed \cite{3}.  \\

Also, by considering non equilibrium systems with a long-term stationary state that possess a spatio-
temporally fluctuating intensive quantity, more general statistics were formulated, called
Superstatistics  \cite{4}.  The temperature was selected as a fluctuating quantity among
various available intensive quantities.  An extensive discussion exists in the literature analyzing 
the possible viability of these kind of models to explain several physical phenomena  \cite{3,5}.  For
general distributions $f(\beta)$, one can get an effective Boltzmann factor.

\begin{equation}
B(E)= \int_0^\infty  d \beta f ( \beta ) \rme^{- \beta E}, \label{1}
\end{equation}
where $E$ is the energy of a microstate associated with each of the considered cells.
The ordinary Boltzmann factor is recovered for $f (\beta) = \delta (\beta-\beta_0)$.  One can, however,
 consider other distributions for the temperature that will lead to their corresponding Boltzmann factors.
 The Gamma $(\chi^2)$, log-normal and the $F$-distributions were studied in this context as well as their
 corresponding Boltzmann factors.  The analysis of these $B(E)$ showed that all these statistics
 present the same behaviour for small variance of the fluctuations.

 In \cite{6} a new formalism was developed to deduce entropies associated to each one of the above
 mentioned Boltzmann factors $B(E)$ arising from their corresponding $f(\beta)$ distributions.  Following
 this procedure, the Boltzmann-Gibbs entropy and the so called non-extensive statistical mechanics,
 Tsallis entropy $S_q$ (corresponding to the Gamma distribution $(\chi^2)$ and depending on a constant parameter $q)$ were
  obtained.  For the log-normal, $F$-distribution and other distributions it is not possible to get
  closed analytic expression for their associated entropies and  the calculations were performed
  numerically utilizing the corresponging $B(E)$ in each case.

  All these $f(\beta)$ distributions and the Boltzmann
  factors $B(E)$ obtained from them, depend then on a constant parameter $q$, actually the $F$-distribution
   also depends  on a second constant parameter.  Consequently the associated entropies depend  on $q$.

In this work we are proposing  new generalized distributions, Boltzmann factors and information entropies depending
  on $p_l$.  It is not our  purpose in this first setting of these new  structures to discuss their possible physical
  consequences or applications.  This is left for future work.  As shown, these new proposals depending on
  $p_l$ resemble the already well known $f(\beta)$, $B(E)$ and information entropies depending on $q$ which possible
     applications and limitations, in relation with certain physical systems, have been  discussed in the literature 
     \cite{3,5,boden}.

  In \Sref{sec2}, we will propose $f(\beta)$ distributions that do not depend on an arbitrary
  constant parameter (the $F$-distribution will also be considered, it will depend now on one constant parameter
  instead of two, as usual), but instead of $p_l$ that can be identified with the
  probability associated with the microscopic configuration of the system.  We will calculate the associated
  Boltzmann factors.  It will be shown that for small variance
  of the fluctuations a universal behaviour is exhibited by these different statistics.

 In \Sref{sec3}, we use the new Gamma $(\chi^2)$ distribution, depending on
  $p_l$, and its associated Boltzmann factor to calculate the entropy; some of these calculations were
   already presented in \cite{7}.  We will maximize this information entropy to get the
   corresponding $\{ p_{l}\}$ probability distribution.  In our model, the entropy resembles
  the one proposed in non-extensive statistical mechanics resulting also by assuming a $(\chi^2)$ distribution.
  In our case, however, it does not depend on a free constant  parameter $q$, but instead on the probability $p_l$.
  We will show that this new entropy can be expanded in a series, which first term corresponds to Shannon  entropy 
   \cite{shannon}.

  In \Sref{sec4}, we will consider other well known information entropies \cite{1}
that can be generalized and the new proposed entropies will not depend on the constant parameter $q$, but on
   the probability $p_l$  in each case. As examples we will
  consider the $p_l$ modified Kaniadakis and Sharma-Mittal  entopies, that, as expected, will have also the
  Shannon entropy as a first term when expanded in series.

  To complete this work we analyze two further aspects, the validity of the saddle-point approximation
  \cite{8, 9, 10},  and following \cite{3,11,12} we also discuss a generalized version of the Khinchin
  axioms.  As shown in \cite{3}, three of these axioms are kept and the fourth of them is replaced by a more
  general version proposed in \cite{11}, obtaining the Tsallis entropy.  We will study these two aspects
   only for the Boltzmann factor $B_{p_l}(E)$  and its associated entropy, arising from the
   proposed $p_l$-Gamma distribution.  For the other entropies we will not discuss these two
   features.  However, by example, for these entropies a similar procedure to that followed in \cite{12}
   could be worked out in relation with an extension of the Khinchin axioms.

   \Sref{sec5} is dedicated to discuss the saddle-point approximation \cite{10}.  We will consider it in relation
   to the Boltzmann factor $B_{p_l} (E)$ arising from the $p_l$-Gamma distribution.  In \Sref{sec6},
   we  discuss how one can replace the fourth Khinchin axiom \cite{3} to get a set of
   axioms from which the entropy proposed here follows.  As mentioned, this entropy is obtained from the
   Boltzmann factor $B p_l (E)$ and this one from the $p_l$-Gamma distribution. \Sref{sec7} is devoted
   to Discussion and Outlook.
   
\section{Generalized distributions and their associated  Boltzmann factors}\label{sec2}

We begin by assuming a Gamma (or $\chi^2$) distributed inverse temperature $\beta$ depending on
$p_l$,   the  probability associated with the microscopic configuration
of the system.  We may write these $p_l$ Gamma distributions as

\begin{equation}
f_{p_{l}}(\beta) = \frac{1}{\beta_0 p_l \Gamma\left(\frac{1}{p_l}\right)}
\left( \frac{\beta}{\beta_0}\frac{1}{p_l}\right)^{\frac{1-p_l}{p_l}}
\rme^{-\beta/\beta_0 p_l}, \label{2}
\end{equation}
where $\beta_0$ is the average inverse temperature.
Integration over $\beta$ yields the generalized Boltzman factor

\begin{equation}
B_{p_{l}}(E) = (1+ p_l \beta_0 E)^{- \frac{1}{p_l}}, \label{3}
\end{equation}
as shown in \cite{4}, this kind of expression can be expanded for small
$p_l\beta_0E$,  to get

\begin{equation}
B_{p_l}(E) = 	\rme^{-\beta_0 E} \left[1+ \frac{1}{2}p_l \beta^2_0 E^2 - \frac{1}{3} p^2_l \beta^3_0 E^3 + ...\right].\label{4}
\end{equation}

The log-normal distribution can also be written in terms of $p_l$ as

\begin{equation}
f_{p_{l}}(\beta) = \frac{1}{\sqrt{2\pi} \beta [ \ln ( p_l+1)]^{1/2}}  \exp
\{ - \frac{ \left[ \ln  \frac{\beta (p_l+1)^{1/2}}{\beta_0} \right]^2}{2 \ln (p_l+1)} \}. \label{5}
\end{equation}
The generalized Boltzmann factor (\ref{1})  can be obtained in leading order, for small variance of the
inverse temperature fluctuations,

\begin{equation}
B_{p_{l}}(E) = \rme^{-\beta_0 E} \left[1 + \frac{1}{2} p_l \beta^2_0 E^2 - \frac{1}{6} p^2_l(p_l+3)\beta^3_0 E^3+ \cdots \right]. \label{6}
\end{equation}

In general, the $F$-distribution has two free constant parameters. In \cite{4}, the authors considered,
particularly, the case in which one of these constant parameters is chosen as $v=4$. For this same
value of this constant parameter we define a $F$-distribution in function of the inverse of the
temperature and $p_l$ as

\begin{equation}
f_{p_{l}}(\beta)= \frac{\Gamma\left(\frac{8p_l-1}{2p_l-1}\right)}{\Gamma\left((\frac{4p_l+1}{2p_-1}\right)}
\frac{1}{\beta_0^2}
\left(\frac{2p_l-1}{p_l+1}\right)^2\frac{\beta} {\left(1+\frac{\beta}{\beta_0}\frac{2p_l-1}{p_l+1}\right)^{\left(
\frac{8p_l-1}{2p_l-1}\right)}}     .  \label{7}
\end{equation}
 Once more the associated Boltzmann factor   can not be evaluated
in a closed form, but for small variance of the fluctuations we obtain the series expansion

\begin{equation}
B_{p_{l}}(E) = \rme^{-\beta_0E} \left[ 1+ \frac{1}{2}p_l \beta^2_0 E^2 + \frac{1}{3} p_l \frac{(5 p_l-1)}{p_l-2} \beta^3_0 E^3 + ... \right]. \label{8}
\end{equation}

Beck and Cohen \cite{4} have demonstrated that all superstatics
depending on a constant parameter $q$ are the same
for sufficiently small variance of the fluctuations.
For
our proposed distributions (\ref{2}, \ref{5}, \ref{7})  depending now on $p_l$,  the
corresponding Boltzmann factors (\ref{4},\ref{6}, \ref{8})  also satisfy these conditions.

\section{Entropy from the Boltzmann factor}\label{sec3}

The examples studied in \cite{4} have been nicely addressed in \cite{6} in order to deduce the entropies  from
their corresponding Boltzmann factors. Another possible way to reconstruct the entropy has been proposed in
\cite{13,14}, this provides other expressions and consequently predicts different physical consequences.
In \cite{15} it has been shown that there exist a duality between these two procedures.  We will refer
to the first proposal, as shown there \cite{6}, the Boltzmann-Gibbs entropy and the non-extensive
statistical mechanics entropy can be obtained in a closed analytic form.  However, the entropies corresponding
to the Boltzmann factors associated to the log-normal and to the $F$-distributions can not be obtained
analitically and were calculated numerically.  Following \cite{6} and a previous work by one of us \cite{7} we present
the procedure to obtain the entropy corresponding to our $f(\beta)$ distribution (\ref{2})  and to its associated
generalized Boltzmann factor (\ref{3}). We begin by defining the entropy $S= \displaystyle\sum_{l=1}^{\Omega} s(p_l)$
in terms of a generic $s(p_l)$; for $s(x)=-x \ln x$ one has the Shannon entropy. As in \cite{6} it is
possible to express $s(x)$ and as well a generic $u(x)$ in terms of integrals on a function $E(y)$ that is
obtained from the Boltzmann factor $B(E)$ of interest.  By these means $s(x)$ and $u(x)$ can be written
as

\begin{eqnarray}
s(x)=\int^x_0  dy \frac{\alpha+E(y)}{1-E(y) /E^\ast}~~, \label{9}
\end{eqnarray}
and
\begin{eqnarray}
u(x)=(1+ \alpha / E^\ast) \int^x_0 \frac{dy}{1-E(y)/E^\ast}~~, \label{10}
\end{eqnarray} 
where $E(y)$ is to be identified with the inverse function of
$B_{p_{l}}(E)/\int^\infty_0  dE'~ B_{p_{l}}(E')$.  One selects first the $f(\beta)$ of interest, then
$B(E)$ is calculated and the integral $\int^\infty_0 B(E')~dE'$ is performed. Inverting the
axes of the variables,  $E(y)$ for several superstatistics   can be found \cite{4}, and from it $E^\ast$.  
In our case, the starting points are the distribution (\ref{2})  
and the  Boltzmann factor (\ref{3}). $E(y)$ results in
\begin{equation}
E(y)= \frac{y^{-x}-1}{x}~~, \label{11}
\end{equation}
$E^\ast = - \frac{1}{x}$. A straightforward calculation gives for
$u(x)$
\begin{equation}
u(x) = x^{x+1} , \label{12}
\end{equation}
where $\alpha$  has been determined by means of the condition $u(1)=1$; $s(x)$ results in
\begin{equation}
s(x) = 1 - x^{x} . \label{13}
\end{equation}
Expressions (\ref{12}, \ref{13})  fulfill the conditions $s(0)= 0, u(0)=0$ and $u(1)=1, s(1) =0$.  By these means the entropy results
in
\begin{equation}
S = k \displaystyle\sum_{l=1}^{\Omega} (1-p_l^{p_l}),\label{14}
\end{equation}
where $k$ is the conventional constant and $\displaystyle\sum_{l=1}^{\Omega} p_l=1$.  The expansion of (\ref{14})  gives

\begin{equation}
-\frac{S}{k}= \displaystyle\sum_{l=1}^{\Omega} p_l \ln{p_l} +
\frac{(p_l \ln{p_l})^2}{2!} + \frac{(p_l
\ln{p_l)}^3}{3!}+\cdots,\label{15}
\end{equation}
the first term corresponds to Shannon entropy.

In the functional
\begin{eqnarray}
\Phi= \frac{S}{k}-\gamma \displaystyle\sum_{l=1}^{\Omega}p_l-\beta \displaystyle\sum_{l=1}^{\Omega}
p_l^{p_{l+1}}E_l,\label{16}
\end{eqnarray}
$\gamma$ and $\beta$ are Lagrange parameters.

\begin{figure}[h]
  \centering
    \includegraphics{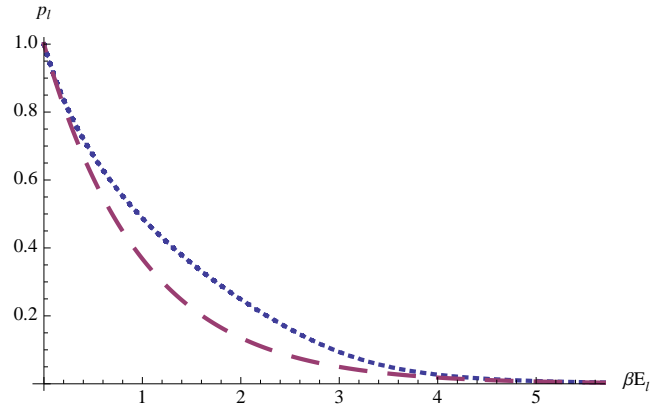}
    \caption{\label{F4} Comparison of the two probabilities.~ \dotted ~line corresponds to the standard one $p_{l}=\rme^{-\beta E_{l}}$, and \broken ~line to $p_{l}=f(\beta E_{l})$}
\end{figure}

Considering the appropiate $s(x)$ and $u(x)$ and imposing
the condition $\frac{\partial \Phi}{\partial p_l}=0$ it is possible to calculate $p_l$, as it is known for the
Shannon entropy \cite{shannon} and for the non  extensive statistical mechanichs \cite{1,6}.  In our case by means of (Eqs. \ref{12},\ref{14})  one
gets
\begin{equation}
1+ \ln p_l + \beta E_l (1+p_l+p_l \ln  p_l) = p_l^{-p_l}, \label{17}
\end{equation}
As we have shown in the expansion   of the entropy   the dominant term is the first
one corresponding to the  Shannon entropy, for it one gets the usual expression
for $p_l$ namely $p_l= \rme^{-\beta_o E_l}$.
We cannot analytically express $p_l$ as function of $\beta E_l$ for the model we are considering.
 We can however, plot a figure of $\beta E_l$ as function of $p_l$ and invert the axes to get
  $p_l = f(\beta E_l)$.   By means of this procedure we are able to plot \Fref{F4} for different
  values of $\beta E_l$.  We notice that for relative large values of $\beta E_l$ the usual values for
  $p_l$ coincide with the ones given by (\ref{17}).  As expected they coincide also for $p_l$
  approaching one.
  
\begin{figure}[h]
    \centering
    \includegraphics{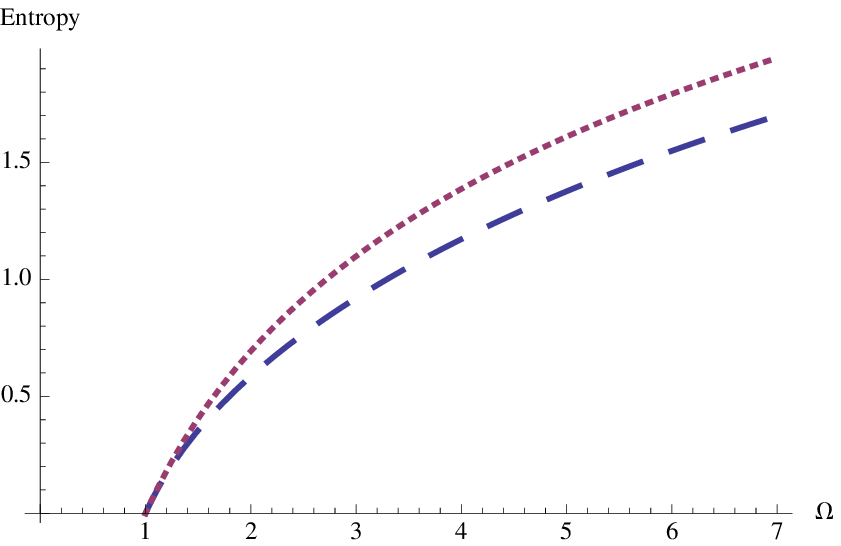}
    \caption{\label{F1} Entropies as function of $\Omega$. \dotted ~ and ~ \broken ~lines correspond to $S_{\beta}/k$ and $S/k$ respectively ($p_{l}=1/\Omega$ equipartition)}
\end{figure}

\begin{figure}[h]
   \centering
    \includegraphics{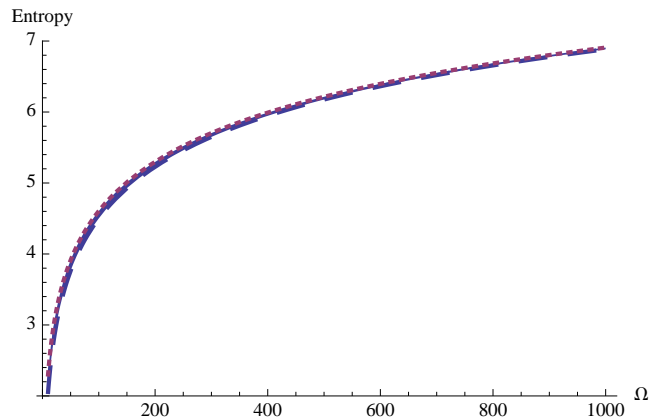}
    \caption{\label{F2} Entropies as function of $\Omega$. \dotted ~and~ \broken ~ lines correspond to $S_{\beta}/k$ and $S/k$ respectively ($p_{l}=1/\Omega$ equipartition)}
\end{figure}

As we have shown by choosing $f_{p_{l}} (\beta)$ (\ref{2}) , $B_{p_{l}} (E)$ (\ref{3})  is obtained by integrating
over $\beta$  and from it by inverting the axes of the variable the inverse function $E(y)$
(\ref{11})  and $E^\star$ are found.  This procedure has allowed us to calculate $u(x)$   and $s(x)$
  and consequently
the entropy (\ref{14},\ref{15}) . If we assume in $f_{p_l} (\beta)$ (\ref{2}) a  $p_l=\frac{1}{\Omega}$, that is
equipartition, the distribution takes the form

\begin{equation}
f_\Omega (\beta) = \frac{\Omega}{\beta_0 \Gamma \left(\Omega\right) }
\left( \frac{\beta \Omega}{\beta_0 }\right)^{\Omega-1}
\rme^{-\frac{\beta\Omega}{\beta_o}}, \label{18}
\end{equation}
and we get the Boltzmann factor 

\begin{equation}
B_\Omega (E) = (1+\beta/\Omega)^{-\Omega}, \label{19}
\end{equation}
from this the entropy results in

\begin{equation}
S=k \Omega \left[ 1- \frac{1}{\Omega^{\frac{1}{\Omega}}} \right], \label{20}
\end{equation}
as Boltzmann's entropy is $S_B=k \ln \Omega$,  the expansion of expression (\ref{20})  in terms of $S_B$ gives

\begin{equation}
\frac{S}{k}= \frac{S_B}{k}- \frac{1}{2 !} \rme^{-S_B/k} \left(\frac{S_B}{k}\right)^2 +  \frac{1}{3 !} e^{-\frac{2S_B}{k}}
 \left(\frac{S_B}{k}\right)^3  \cdots   . \label{21}
\end{equation}
Figures \ref{F1} and \ref{F2} show the Boltzman entropy $S_\beta/k$ and the entropy 
$S/k$.  We notice that in the range of "small" $\Omega$
(large equipartition probabilities) these entropies differ.  Instead for large $\Omega$ the
two entopies basically coincide.

We have then proposed a new entropy (\ref{14},\ref{20})  that does not depend on a constant arbitrary parameter, but
on the probability $p_l$  (\ref{17}) associated with the
microscopic configuration of the system. Its expansion provides
 as a first term the Shannon entropy (\ref{15})  and correspondingly Boltzmann's entropy (\ref{21}).
This entropy corresponds to the Gamma distribution (\ref{2},\ref{18}).  Other
two distributions (\ref{5},\ref{7})  were also assumed as functions of $p_l$ and their approximated 
corresponding Boltzmann factors found (\ref{6},\ref{8}).  As mentioned, their associated entropies can not be expressed in an analytic
closed form.

\section{Other generalized information entropies in terms of $p_l$  }\label{sec4}

More general measures of information than the Shannon entropy have been proposed in the literature
\cite{1}.  Maximizing these entropies subject to suitable constraints allow us to obtain
associated probability distributions. In \cite{3,5,boden} several of these entropies have been reviewed and their
 potential physical applications discussed.

Similar entropies, to most of those studied in \cite{3}, but now in terms of  $p_l$, can be proposed.
 As examples let us consider modified
 Kaniadakis and Sharma-Mittal entropies.  The Kaniadakis entropy is defined by the expression

\begin{equation}
 S_\kappa = - k \displaystyle\sum_{l}^{\Omega} \frac{p_l^{1+\kappa}-p_l^{1-\kappa}}{2\kappa}. \label{22}
\end{equation}
This is an entropy, which reduces to the original Shannon entropy  for $\kappa=0$ \cite{3}.
 Inspired in this Kaniadakis entropy we propose to consider the following generalized entropy

\begin{equation}
S=- k \displaystyle\sum_{l}^{\Omega} \frac{p_l^{p_l}-p_l^{-p_l}}{2}, \label{23}
\end{equation}
the two terms in this expression can be expanded in a similar manner as (\ref{14},\ref{15})  to get

\begin{equation}
-\frac{S}{k}=  \displaystyle\sum_{l}^{\Omega} p_l \ln p_l + \frac{(p_l \ln{p_l})^3}{3!} + \cdots .  \label{24}
\end{equation}
The first term corresponds to Shannon entropy. It is interesting to notice that  the expansion
(\ref{24})  of the entropy (\ref{23})  differs from the expansion (\ref{15})  of the entropy (\ref{14})  
corresponding to the Gamma distribution (\ref{2})  and that we have analyzed in more detail in the previous sections;
in (\ref{24}) only the "odd" terms in the expansion arise.

We consider now the Sharma-Mittal entropies, these are two constant parameters families of entropic
forms.  They can be written as

\begin{equation}
S_{\kappa,r} =-k \displaystyle\sum_{l}^{\Omega} p_l^r \left( \frac{p_l^\kappa -p_l^{- \kappa}}{2\kappa}  \right). \label{25}
\end{equation}
We now assume that $\kappa$ and $r$  are not constant parameters but are
functions only of $p_l$, we get the entropy (\ref{14})  for $-2\kappa=p_l$ and $r=\frac{p_l}{2}+1$ and the entropy 
(\ref{23})
is obtained for $r=1$ and $\kappa=p_l$.  These Sharma-Mittal entropies can be generalized in several manners as functions
of the  probability $p_l$ by means of other different assumptions.  Two of them correspond to the entropies
(\ref{14},\ref{23})
in this work. Other entropies considered, by example in \cite{3}, can also  be generalized as functions of
 $p_l$ instead of a constant parameter $q$.  As examples we have analyzed the $p_l$ dependent generalized
 Kaniadakis and two Sharma-Mittal entropies that reduce to the entropies (\ref{14},\ref{23}).

\section{The saddle-point approximation, the $p_l$-Gamma distribution case.}\label{sec5}

We have already obtained the low energy asymptotics (\ref{4}) for the Boltzmann factor (\ref{3})
 arising from the $p_l$-Gamma distribution (\ref{2}).  We have also shown that up to second order
 the Bolztmann factor (\ref{6}) corresponding to the log-normal distribution (\ref{5}) and the
  Boltzmann factor (\ref{8}) arising from the $F$-distribution (\ref{7}) concide with the
  Boltzmann factor (\ref{4}).  These approximations represent the leading order correction to ordinary
  statistical mechanics in the nonhomogeneous  systems with temperature fluctuations for small values
  of the energy $E$.  The zeroth-order approximation  to these Boltzmann factors corresponds, as
  expected, to the Boltzmann statistics $B(E) \sim \rme^{\beta_0E}$ with inverse temperature $\beta_0$.

  To find the high-energy asymptotics of $B(E)$ we follow \cite{10} where the fact is used that
  (\ref{1}) has the form of a Laplace integral for $E \rightarrow \infty$.  In this limit, the integral
   can be approximated by its largest integrand.  This is the essence of the saddle-point approximation,
   namely the Laplace method.  The conditions of applicability of this approximation method are basically the
   conditions that one assumes regarding the shape of $f(\beta)$ and its differenttiability.  By putting
   $B_{p_l} (E)$ in the form

\begin{equation}
B_{p_l}(E) = \int^\infty_0 \rme^{-\beta E+ \ln f(\beta)}d\beta, \label{26}
\end{equation}
one attempts to find the unique value of $\beta$  which maximizes the exponential function

\begin{equation}
\Psi(\beta,E)=-\beta E+ \ln f(\beta) , \label{27}
\end{equation}
for any  large enough energy value $E$.  The value of $\beta$ maximizing $\Psi(\beta,E)$ for large
 fixed $E$ is denoted $\beta_E$.  Having assumed $f(\beta)$ to be unimodal ensures us the uniqueness
 of $\beta_E$.
Being $f(\beta)$ unimodal, $\ln f(\beta)$ must be a concave function of $\beta$.  In this
manner the maximum of $\Psi(\beta,E)$ can only be obtained at the single point $\beta_E$. It
is such that

\begin{equation}
E=[\ln f_{p_l}(\beta)]^\prime = \frac{f_{p_l}^\prime (\beta)}{f_{p_l}(\beta)} . \label{28}
\end{equation}
In this way we get $\beta_E$ and in the limit $E\rightarrow \infty$

\begin{equation}
B_{p_l}(E)\sim \rme^{\Psi(\beta_E,E)} = f_{p_l}(\beta_E)\rme^{-\beta_E E}. \label{29}
\end{equation}
This saddle-point or Laplace approximation can be improved by using a Gaussian approximation of the integrand
in (\ref{26}). The refined high energy asymptotics results in

\begin{equation}
B_{p_l}(E)\sim \frac{f_{p_l}(\beta_E)\rme^{-\beta_E E}}{{\sqrt {- [\ln f_{p_l} (\beta_E)]^{\prime \prime}} }}. \label{30}
\end{equation}

The approximation of $B(E)$ (\ref{29}, \ref{30}) show that the mixture of Boltzmann statistics
defining $B(E)$ reduces at high energy $E$ to a particular Boltzmann statistics, like in the equilibrium
situation, but now this Boltzmann statistics is a function of $\beta_E$ which depends on $E$, the energy
considered  and is determined by $f_{p_l}(\beta)$ (\ref{28}).  The long term stationary behaviour of the non
equilibrium system considered for high values of $E$  is dominated by the equilibrium behaviour of a
subset of cells having an inverse temperature close to $\beta_E$.
We now consider the asymptotic behaviour of $B_{p_l}(E)$ (\ref{1}) for $E\rightarrow \infty$  and
$f_{p_l} (\beta)$  given by (\ref{2}).  We  first solve (\ref{28}) to find $\beta_E$ for this case.
One gets

\begin{equation}
\beta_E = \frac{(1-p_l)\beta_0}{E p_l \beta_0 +1}, \label{31}
\end{equation}
as expected \cite{10} as $E\rightarrow \infty$, $\beta_E \rightarrow 0$.
We want now to calculate $B_{p_l}(E)$ (\ref{29})  for this $\beta_E$.  This can be expressed as

\begin{equation}
B_{p_l} (E) = \frac{1}{\beta_0 p_l \Gamma (\frac{1}{p_l})} \rme^{-\frac{1-p_l}{p_l} [\ln \frac{p_l}{1-p_l} + \ln
(E p_l \beta_0 +1) +1]} , \label{32}
\end{equation}
For $E\rightarrow \infty$ and a certain value of $p_l$

\begin{equation}
B_{p_l} (E) \sim \rme^{- \frac{1-p_l}{p_l} \ln E}  \sim  E ^{1- \frac{1}{p_l}}  . \label{33}
\end{equation}
The more refined approximation (\ref{30})  can be obtained by dividing (\ref{33}) by
 $\sqrt{- [\ln f_{p_l}(\beta_E)]^{\prime\prime}}$ which in the high energy limit is proportional to $E$.
 In this way

\begin{equation}
B_{p_{l}}(E) \sim E^{- 1/p_l}. \label{34}
\end{equation}
In this more accurate  calculation, we get, asymptotically a decaying power law for the effective
Boltzmann factor.  As mentioned in \cite{10} power law superstatistics seem to be physically relevant
 for several physical systems.

\section{An appropiate information measure}\label{sec6}

The well established four Khinchin axioms are nicely well discussed and presented in \cite{3}. 
As known, the celebrated Shannon \cite{shannon} entropy $S=-k  \displaystyle\sum_{l}^{\Omega} p_l \ln p_l$ satisfies all these axioms.
It has been however, argued in the literature that the fourth of these axioms is not an obvious property \cite{3,11,12}.
We will concentrate our discussion on it.  This fourth axiom deals with the composition of two systems
I and II (not necessarily independent).  We denote the probabilities of the first system as $p^I_i$, those
of the second system as $p^{II}_{j}$.  The joint system is described by the joint probabilities $p^{I,II}_{ij} = p^I_i
p^{II}(j|i)$, where $p^{II}( j|i)$ is the conditional probability of event $j$ in system II under the condition
that event $i$ has already ocurred in system I.  The conditional information of system II formed with
the conditional probabilities $p^{II} (j|i)$ is denoted by $I( \{ p^{II} (j|i)\} )$, under the condition that
system I is in the state $i$.  The fourth axiom states that the conditional informations are related by

\begin{equation}
I( \{ p^{I,II}_{ij} \})=I( \{ p^I_i \}) + \displaystyle\sum_{i} p^I_i I ( \{ p^{II} (j|i) \}) . \label{35}
\end{equation}
This axiom postulates that the information measure should be independent of the way the
information is collected.  We can collect the information in II, assuming a given
event $i$ in system I, and then sum the result over all possible events $i$ in  system I,
weighting with the probabilities $p^I_i$.  If the two systems are independent the probability of the
two systems factorizes $p^{I,II}_{ij}= p^I_i p^{II}_j$.  Only in this case (\ref{35}) reduces
to

\begin{equation}
I(\{ p^{I,II}_{ij})=I(\{p^I_i\}) + I (\{p^{II}_j\}),  \label{36}
\end{equation}
the rule of additivity of information for independent systems.  From a physical point of view this
axiom (\ref{35}) is not an obvious property.  Should the information be considered as
independent from the way we collect it?  In complex systems, the order in which the information
is collected can be very relevant.  This has lead to the replacement of the fourth Khinchin axiom
by  something more general.  In particular in \cite{11} it was shown that Tsallis entropy
follows uniquely by replacing only the fourth axiom (\ref{35}) by the more general version

\begin{equation}
S^{I,II}_q = S^I_q +S^{II |I}_q - (q-1) S^I_q S^{II|I}_q . \label{37}
\end{equation}
The meaning of this new axiom is that if we collect information from two subsystems, the
total information should be the sum of the information collected from system I and the conditional
information from system II, plus a correction term.  Apriori this correction term can be anything.
One restricts the possible asumptions to

\begin{equation}
S^{I,II}= S^I + S^{II|I} + g (S^I, S^{II|I}), \label{38}
\end{equation}
where $g$ is some function.  One of the simplest forms is of the kind given by (\ref{37}).
We may as well, formulate another axioms which then lead to other possible information measures.
This is the case, by example, in \cite{12} where a set of axioms has been assumed
that lead to the Sharma-Mittal entropy.

The entropies (\ref{14}) associated with the $p_l$-Gamma distribution (\ref{2}) are composable.
Suppose the two systems I and II are not independent.  In this case one can still write the joint
probability $p_{ij}$ as a product of $p_i$ and the conditional probability $p(j|i)$, the probability
of event $j$ under the condition that event $i$ has already ocurred is $p_{ij}=p(j|i) p_i$.

Then the conditional entropy associated with system II, under the condition that system I is in
state $i$, is $(k=1)$

\begin{equation}
S^{II|i}_{p_{j|i}} = 1 - \displaystyle\sum_{j} p^{p^{(j|i)}}_{(j|i)}. \label{39}
\end{equation}
One can then verify the condition

\begin{equation}
S^I_{p_i} + \displaystyle\sum_{i} p^{I~p^{I}_{i}}_i S^{II|i}_{p(j|i)} = S^{I,II}_{p_{ij}}.   \label{40}
\end{equation}
This relation is similar to the original axiom four (\ref{35}), one has however
the probability with an exponent that is the probability itself.  We weight now the events
in system I with $p^{p_i}_i$ instead of $p_i$.  Hence, the $p_l$ dependent information
 considered (\ref{14}) is not independent of the way it is collected for the various
 subsystems.

\section{Discussion and Outlook}\label{sec7}

The distributions (\ref{2},\ref{5},\ref{7}), Boltzmann factors (\ref{3},\ref{4},\ref{6},\ref{8})
and entropies  (\ref{14},\ref{20},\ref{23},\ref{25})  proposed
in this work do not depend on a constant parameter $q$, but all are functions of  $p_l$ (for
the $F$-distribution the remaining constant parameter has been chosen $v=4$). By maximizing the functional
 (\ref{16}), for the entropy (\ref{14})  a closed relation between
$p_l$ and $\beta E_l$ has been obtained (\ref{17}). It has been shown that for small variance of the
fluctuations,  the Boltzmann factors (\ref{4},\ref{6},\ref{8}) concide for the following distributions:
Gamma (\ref{2}) , log-normal (\ref{5})  and $F$ -distributions  (\ref{7}). Moreover,
the generalized information entropies proposed in this work (\ref{14},\ref{23},\ref{25})  can be expanded to get as
a first term the Shannon entropy (\ref{15},\ref{24}).  We have analyzed the saddle point approximation
for the $p_l$-Gamma distribution and have got, in the high energy limit, an asymtotically decaying
power law for the effective Boltzmann factor (\ref{34}), this seems to be the physically expected
 appropiate behaviour \cite{10}.  We have also shown how the fourth Khinchin axiom should be modified
 so that the associated entropy results in  (\ref{14}).  In figure 1 we have compared the probability distribution
 arising from this entropy with the standard one.  In figures 2 and 3 we have compared (for $p_l=\frac{l}{\Omega}$, 
 equipartition)
 the entropy (\ref{20}) with the Boltzmann entropy.  In this first proposal of these new entropies we do not analyze 
 physical
systems in which these entropies could possibly be of interest.  As shown, the new entropies (\ref{14},\ref{23},\ref{25})
 resemble the well known non-extensive statistical mechanichs entropy, the Kaniadakis and two Sharma-Mittal
 entropies correspondingly (other well known $q$ entropies could also be generalized in terms of  $p_l$).
 For all these, depending on a constant parameter $q$,   there exist an extensive literature on their
   possible physical reach,  their  relation with the experiments is under discussion \cite{3,5,boden}. Also, other 
   theoretical developments have been proposed  \cite{13,14,15}.  We will
  study, in further work, some of the $p_l$ dependent entropies proposed here in connection with these aspects.

\ack{Thanks are expressed to Alberto Robledo for explaining us some aspects of generalized information entropies and
for guiding us to literature.  The proposals in this work are, however, entirely our responsability.
S. Zacar\'{i}as helped with the elaboration of the  figures
and is also acknowledge.
This work was supported by CONACYT project 135026, PROMEP and UG projects.
}

\end{document}